\documentclass{Interspeech}

\interspeechcameraready

\title{MPE-TTS: Customized Emotion Zero-Shot Text-To-Speech Using Multi-Modal Prompt}

\author[affiliation={1}]{Zhichao}{Wu}
\author[affiliation={2}]{Yueteng}{Kang}
\author[affiliation={2}]{Songjun}{Cao}
\author[affiliation={2}]{Long}{Ma}
\author[affiliation={1}]{Qiulin}{Li}
\author[affiliation={1}]{Qun}{Yang}

\affiliation{}{Nanjing University  of Aeronautics and Astronautics}{China}
\affiliation{Youtu Lab}{Tencent}{China}
\email{zhichao\_wu@foxmail.com, kangyueteng@foxmail.com, xmdxcsj@gmail.com, malong@gmail.com, qiulin\_li@nuaa.edu.cn, qun.yang@nuaa.edu.cn}

\usepackage{comment,float}

\begin{document}

\maketitle

\begin{abstract}

Most existing Zero-Shot Text-To-Speech(ZS-TTS) systems generate the unseen speech based on single prompt, such as reference speech or text descriptions, which limits their flexibility. We propose a customized emotion ZS-TTS system based on multi-modal prompt. The system disentangles speech into the content, timbre, emotion and prosody, allowing emotion prompts to be provided as text, image or speech. To extract emotion information from different prompts, we propose a multi-modal prompt emotion encoder. Additionally, we introduce an prosody predictor to fit the distribution of prosody and propose an emotion consistency loss to preserve emotion information in the predicted prosody. A diffusion-based acoustic model is employed to generate the target mel-spectrogram. Both objective and subjective experiments demonstrate that our system outperforms existing systems in terms of naturalness and similarity. The samples are available at \url{https://mpetts-demo.github.io/mpetts_demo/}.


\end{abstract}

\section{Introduction}

The conventional TTS systems\cite{fastspeech2,gradtts,tacotron2} are limited because they only generate the speech of seen speakers. In order to meet practical application scenarios, Zero-Shot TTS(ZS-TTS) which aims to generate speech with unseen style has been introduced. Generally, ZS-TTS systems face challenges in generalization, fine-grained customization and modeling speech nuances. In recent years, research on ZS-TTS models has primarily focused on two main areas: speech-based ZS-TTS and text-based ZS-TTS.

Speech-based ZS-TTS is predominantly employed for generating speech of unseen speakers, as speech clips encompass both semantic and acoustic information, enabling the model to capture speaker-specific features. Most of these methods \cite{megatts,valle,gst, meta,generspeech} extract the target style features from the reference speech clips and keep the style consistency. For example, Global style token (GST) \cite{gst} designs a style token layer and a reference encoder to achieve the expressive TTS systems. Meta-StyleSpeech\cite{meta} adopts the base architecture upon FastSpeech2\cite{fastspeech2}, applying style adaptive layer norm and meta-learning algorithm to effectively synthesize style-transferred speech. GenerSpeech\cite{generspeech} proposes a multi-level style adapter and a generalizable content adapter to efficiently model the style information. These methods model the reference speech clip as the style feature, incorporating various speech attributes such as timbre, emotion and prosody. However, the entanglement of features makes fine-grained customization extremely challenging, which in turn hinders the achievement of style transfer and the control of emotional expression in speech generation. Both of these capabilities are crucial for effective ZS-TTS.

Recently, the advent of large language models and the development of text-based image generation have significantly increased interest in using text descriptions as prompts to guide speech generation. A number of studies\cite{prompttts,instructtts,prompttts2} have proposed the use of prompt encoders to extract the style representations from the text prompts, which then guide zero-shot speech generation. For instance, PromptTTS\cite{prompttts} incorporates a style encoder that maps a text prompt to a semantic space to extract the style representation, thereby guiding the content encoder and speech decoder. InstructTTS\cite{instructtts} effectively captures semantic information from text prompts and models acoustic features in a discrete latent space. Furthermore, InstructTTS employs a discrete diffusion model to generate discrete acoustic features. Despite the ability of text-based ZS-TTS systems to generate speech with specific styles, users often struggle to formulate accurate natural language descriptions for their desired speech style. To address this issue, multi-modal support has emerged as a viable solution, offering users a broader range of options for guiding speech generation.
\begin{figure*}[h]
	
	\begin{minipage}{0.5\linewidth}
		\vspace{3pt}
		\centerline{\includegraphics[width=240pt]{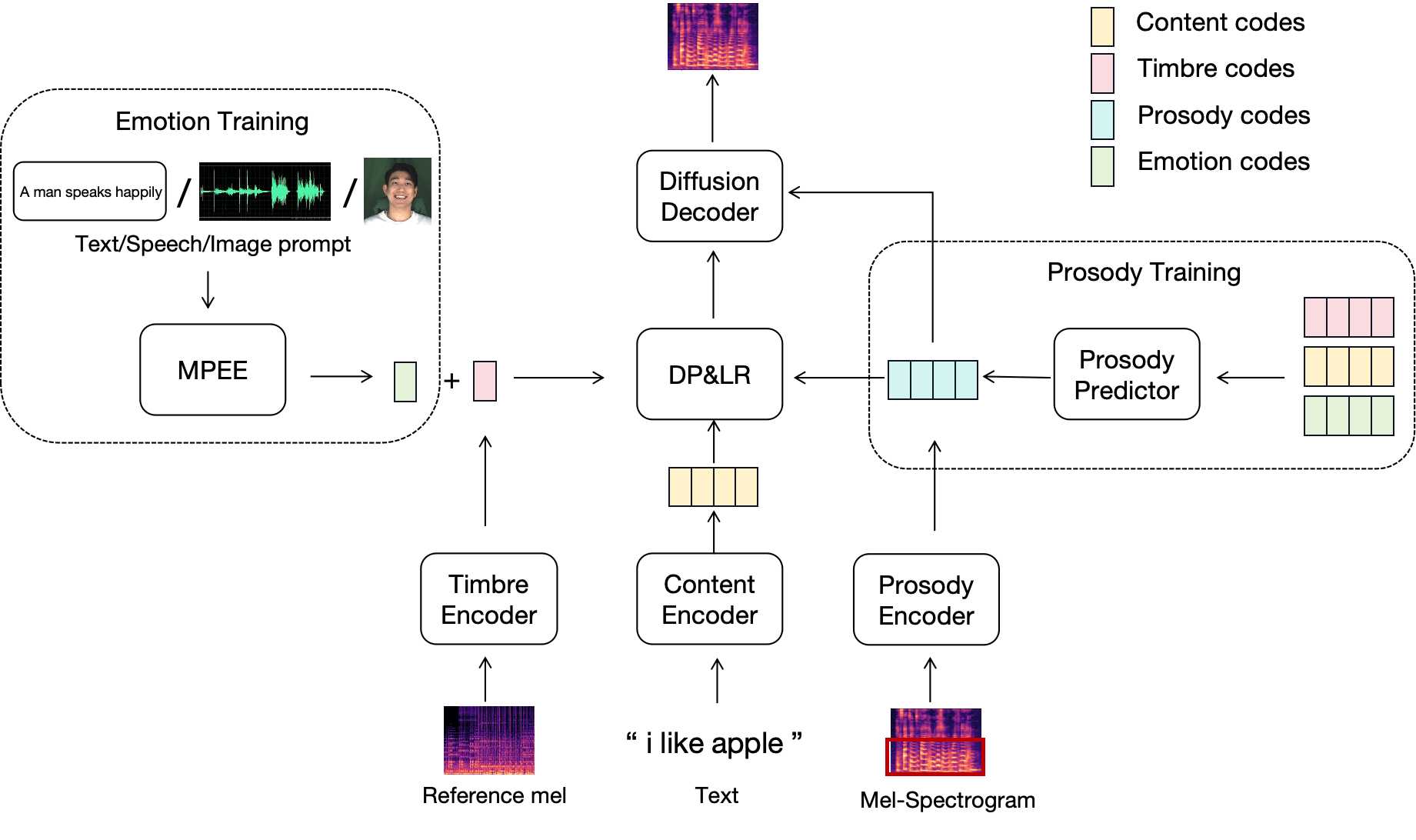}}
		\centerline{(a)}
	\end{minipage}
	\begin{minipage}{0.5\linewidth}
		\vspace{3pt}
		\centerline{\includegraphics[width=220pt]{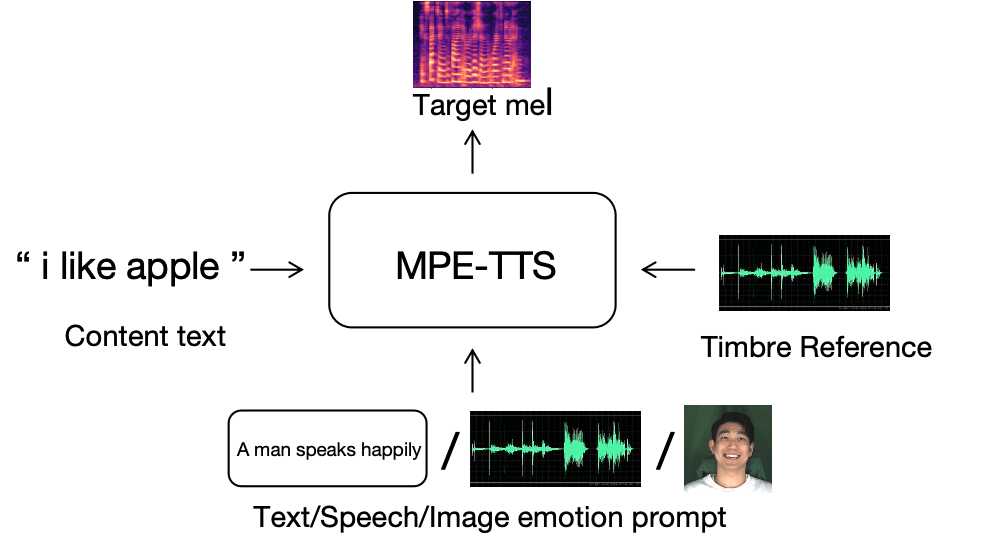}}
		\centerline{(b)}
	\end{minipage}
	\caption{(a) shows the overview of the proposed method. Our framework includes three training stages, in sequence: 1) Emotion Training. 2) Acoustic Model Training. 3) Prosody Training. (b) shows the inference phase of the proposed system }
	\label{fig_model}
\end{figure*}

To the best of our knowledge, MM-TTS\cite{MMTTS} is currently the only system that employs multi-modal prompts to guide ZS-TTS. MM-TTS introduces an Aligned Multimodal Prompt Encoder (AMPE) to unify prompts from different modalities into a stylistic latent space, which then guides the acoustic model to generate mel-spectrograms. Besides, it proposes a Refiner based on Rectified Flow\cite{flow} to refine the generated mel-spectrograms, addressing the problem of over-smoothing. Nevertheless, MM-TTS still exhibits limited capability in zero-shot speech generation.

Inspired by MM-TTS, this paper proposes the MPE-TTS system that supports multiple modal prompts(text, image and speech). Our approach allows users to select one of three modalities as the emotion prompt, which is particularly useful when users lack image or reference speech data, or find it difficult to formulate appropriate text descriptions. Our approach not only provides users with the flexibility to choose their preferred prompt but also enables fine-grained customization over emotions and timbre by the proposed disentangling strategy. Additionally, we employ an LLM-like prosody predictor to capture spontaneous speech nuances, boosting prosody prediction accuracy. The contributions of this paper are as follows:

1) We propose a hierarchical disentangling strategy, modeling speech features at different granularity levels, and achieve effective disentangling.

2) We propose a multi-modal prompt emotion encoder based on Emotion2Vec\cite{emotion2vec} to extract emotion information from text, image or speech prompt. It enables users to flexibly choose one of the three prompt modals.

3) We introduce an LLM-like prosody predictor to predict the target prosody which bring natural prosody performance. Besides, we propose the emotion consistency loss(ECL) to enhance the emotion information in the predicted prosody.

4) Subjective and objective experiments have shown that our system has the ability of unseen generation and achieves flexible customization.

\section{Method}

As depicted in Figure \ref{fig_model}(a), MPE-TTS is composed of three main components: the multi-modal prompt emotion encoder(MPEE), the diffusion-based acoustic model and the LLM-like prosody predictor. These components meticulously model different speech attributes, achieving the disentanglement of speech attributes. Firstly, MPEE extracts the emotion codes from arbitrary emotion prompt. Secondly, we employ diffusion-based acoustic model to generate the target mel-spectrogram, which consists of timbre encoder, content encoder, prosody encoder, duration predictor(DP), length regulator(LR) and diffusion decoder. Moreover, we train an LLM-like prosody predictor to fit the distribution of prosody based on content, emotion and timbre, leveraging the ability of LLMs to capture both local and long-range dependencies. During inference, the target speech is generated by integrating the content from the given text sequence, the timbre extracted from the reference speech, and the prosody predicted by the prosody predictor. As illustrated in the Figure \ref{fig_model} (b), MPE-TTS has the capability to generate arbitrary timbre and emotion during inference, without being limited to a single reference speech. 

\subsection{Disentangling strategy}
In speech signals, information such as content, emotion, timbre, and prosody are integrated, forming a complete speech signal. To achieve fine-grained customization(customizing the timbre, emotion or prosody separately), it is necessary to disentangle these coupled speech features into independent ones. Based on the distinct characteristics of these features, they are hierarchically categorized into coarse-grained and fine-grained features:

1) Coarse-grained features: we define the timbre and emotion features as coarse-grained features because, in general, a speaker’s timbre and emotion do not change significantly within a segment of speech. Modeling these features as global vectors is more efficient and practical.

2) Fine-grained features: we define the content and prosody features as fine-grained features. Content information is frame-related and temporal, requiring fine-grained modeling to ensure semantic accuracy. Since prosody information has a high dynamic range, it is also modeled at the frame level to simulate the fluency and naturalness of real human speech.

After determining the granularity of different features, we adopt finely designed bottlenecks\cite{autovc} for information filtering and cleverly design the inputs of various encoders to help disentangling. Specifically, we extract the global speaker vector from a random sentence of the same speaker using the timbre encoder to disentangle the timbre and content information\cite{autovc,split}. We introduce a VQ-based prosody encoder that extracts ground truth prosody features based on the low 20 bins of each ground truth mel-spectrogram which contain almost complete prosody and significantly less timbre and content information compared to the full band\cite{prosospeech}. During the training of the prosody predictor, we use the emotion feature exacted from a random speech clip with the same emotion label.

\subsection{Multi-Modal Prompt Emotion Encoder}
\label{ssec:emotion_encoder}
In order to achieve more flexible control over the emotion of generated speech, it is necessary to extract emotion features from the prompt of any input modality: text, image and speech prompt. We introduce the pretrained Emotion2Vec as the speech emotion encoder, which serves as a robust and generalizable foundation model for emotion features. For the text and image encoding, inspired by the MM-TTS\cite{MMTTS}, we add a learnable adapter layer subsequent to the fixed CLIP\cite{clip} encoder. To unify the text, image, and speech modalities into a unified emotion latent space, we employ the Mean Squared Error (MSE) loss function to train the text emotion encoder and the image emotion encoder. The total MPEE loss is computed as follows:

\begin{figure}
    \centering
    \includegraphics[width=7cm]{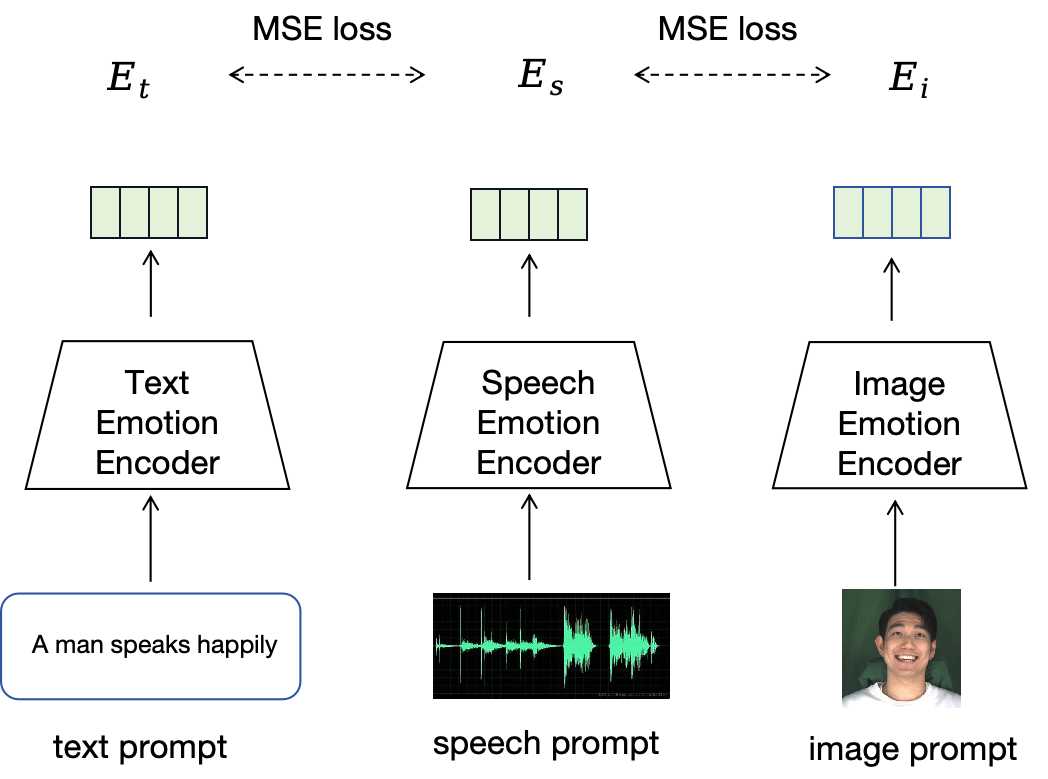}
    \caption{the multi-model prompt emotion encoder }
    \label{Figure:promptencoder}
\end{figure}
\begin{equation} 
Loss_{MPEE}=MSE(E_t,E_s)+MSE(E_i,E_s)
\end{equation}

where ${E_t}$,${E_i}$ and $E_s$  denote the emotion codes extracted from the text prompt, the emotion codes from the image prompt and the emotion codes from the speech prompt respectively.  
During inference, users can choose any modality of emotion prompt to obtain the final emotion codes.

\subsection{LLM-like Prosody Predictor}
\label{ssec:prosody_predictor}
 The Auto-regressive(AR) Transformer\cite{transformer}, which plays a crucial role in large language models, has also been shown to be effective in speech modeling\cite{megatts,tortoise}. Given its excellent capabilities in capturing both local and long-range dependency, we introduce an LLM-like model as the prosody predictor with the aim of modeling the target prosody based on content, timbre, and emotion. 

To train the prosody predictor, we introduce a VQ-based prosody encoder that extracts ground truth prosody codes from the low 20 bins of each ground truth mel-spectrogram. These bins contain almost complete prosody and significantly less timbre and content information compared to the full band\cite{prosospeech}. Note here, in order to disentangle the emotion from other speech features as much as possible, the emotion codes  are extracted from selected speech clips which have similar emotion label but different speaker and content. The prosody predictor is trained in the teacher-forcing mode via the cross-entropy loss. 

To preserve the emotion information in the predicted prosody, we propose the Emotion Consistency Loss(ECL). Specifically, we add a classifier subsequent to the prosody predictor to recognize emotions from the predicted prosody codes, aiming to achieve consistency between the predicted results and the emotion labels. The ECL is computed using cross-entropy loss.

\subsection{Diffusion-Based Acoustic Model}
\label{ssec:diffusion}
The AM backbone model consists of a content encoder, a timbre encoder, a prosody encoder, a duration predictor(DP), a length regulator(LR) and a diffusion decoder. A conformer-based model is used as the content encoder due to its superior performance in capturing local information, thereby achieving more accurate content information extraction. The duration predictor combines content, timbre and prosody to predict the phoneme duration, resulting in more accurate and natural duration. The length regulator upsamples the phoneme-level latent to mel-level latent. The diffusion-based\cite{gradtts} decoder contributes to the high quality of the generated speech.

We first train the acoustic model(AM), employing duration loss, VQ loss, and diffusion loss, which are similar to those used in VQGAN\cite{vqgan} and Grad-TTS\cite{gradtts}. Upon completion of the AM training, we utilize the trained prosody encoder to generate the target prosody features for the training of the prosody predictor.

\begin{table*}[!h]
\centering
\caption{The comparison of Zero-Shot generation based on speech prompt}
\begin{tabular}{l|l|l|l|l|l}
\hline
Model        & MOS $\uparrow$ & ESMOS $\uparrow$ & SSMOS $\uparrow$  & WER $\downarrow$ & ACC $\uparrow$ \\ \hline
GT(mel)      & $4.05\pm0.11$  &   $-$   & $-$ & 18.8 & 54\% \\
Meta-StyleSpeech  & $3.43\pm0.05$  & $3.45\pm0.06$ & $3.53\pm0.06$  &26.0 & 24\% \\
GenerSpeech & $3.53\pm0.06$  & $3.63\pm0.06$ & $3.65\pm0.05$ & \textbf{22.8} & 38\%\\
MM-TTS    & $3.55\pm0.07$  &  $3.75\pm0.06$  & ${3.55\pm0.05}$  & 31.2  & 41\%\\ \hline
Ours       & \bm{ $3.73\pm0.05$ } &  \bm{$4.05\pm0.06$}  & \bm{$3.73\pm0.06$}  &23.4 & \textbf{48\%} \\ 
Ours w/o MPEE & $ 3.20\pm0.06$ &  $ 3.35\pm0.06$ & $3.63\pm0.05 $  & 29.7 &  35\% \\
Ours  w/o ECL & $3.53\pm0.05 $ &  $3.83\pm0.06 $  & $ 3.65\pm0.06$  & 25.4 & 45\% \\ \hline
\end{tabular}
\label{tabel:Table1}
\end{table*}

\begin{table*}[!h]
\centering
\caption{The comparison of Zero-Shot generation based on text prompt.}
\begin{tabular}{l|l|l|l|l|l}
\hline
Model        & MOS $\uparrow$ & ESMOS $\uparrow$ & SSMOS $\uparrow$  & WER $\downarrow$ & ACC $\uparrow$ \\ \hline
GT(mel)      & $4.05\pm0.11$  &   $-$   & $-$ & 18.8 & 54\% \\
MM-TTS    & $3.53\pm0.05$  &  $3.63\pm0.05$  & ${3.53\pm0.05}$  & 32.6  & 37\%\\ \hline
Ours       & \bm{$3.65\pm0.05$} &  \bm{$3.85\pm0.05$} & \bm{$3.73\pm0.05$}  &\textbf{24.5} & \textbf{46\%} \\
Ours w/o MPEE       & $3.45\pm0.05$ &  $3.33\pm0.05$  & $3.45\pm0.05$  &29.7 & 32\%\\
Ours w/o ECL  & $3.55\pm0.05$ &  $3.73\pm0.05$  & $3.63\pm0.05$  &25.3 & 36\%\\ \hline
\end{tabular}
\label{tabel:Table2}
\end{table*}

\begin{table*}[!h]
\centering
\caption{The comparison of  Zero-Shot generation  based on image prompt.}
\begin{tabular}{l|l|l|l|l|l}
\hline
Model        & MOS $\uparrow$ & ESMOS $\uparrow$ & SSMOS $\uparrow$  & WER $\downarrow$ & ACC $\uparrow$ \\ \hline
GT(mel)      & $4.05\pm0.11$  &   $-$   & $-$ & 18.8 & 54\% \\
MM-TTS    & $3.55\pm0.07$  &  $3.53\pm0.05$  & $3.45\pm0.05$  & 31.7  & 35\%\\ \hline
Ours       & \bm{$3.63\pm0.05$} &  \bm{$3.90\pm0.04$} & \bm{$3.76\pm0.05$}  &\textbf{23.4} & \textbf{47\%} \\
Ours w/o MPEE       & $3.43\pm0.05$ &  $3.45\pm0.05$  & $3.55\pm0.07$  &27.7 & 29\%\\
Ours w/o ECL  & $3.55\pm0.06$ &  $3.75\pm0.06$  & $3.63\pm0.05$  &24.3 & 34\%\\ \hline
\end{tabular}
\label{tabel:Table3}
\end{table*}

\section{Dataset and Experiments}
\label{sec:pagestyle}

\subsection{Construct the Multi-modal Emotion Dataset}
\label{sec:dataset}
During the training, we utilized the LibriTTS\cite{libritts} and MEAD-TTS\cite{MMTTS} datasets. The LibriTTS dataset, a multi-speaker English corpus of approximately 585 hours of read English speech at 24kHz sampling rate, was used for the pretraining of MPE-TTS, enabling it to acquire fundamental generation capabilities. MEAD-TTS is a multi-modal TTS corpus featuring 48 actors and actresses that talk with 8 different emotions at 3 different intensity levels. Specifically, it consists of 31055 triples of (speech, transcript, face image) data with a total duration of approximately 36 hours, and the sample rate is 16khz. We selected 6 speakers as the test set. Given the absence of text prompt in the MEAD-TTS data, we generated them by ourselves. Similar to other works\cite{MMTTS,tango2}, we used a large language model to generate text prompts, thereby constructing a multi-modal prompt dataset of text, image, speech. Specifically, we requested language model\footnote{https://github.com/QwenLM/Qwen} to generate synonym text descriptions  based on the original labels (gender, emotion, emotion intensity) in the MEAD-TTS. This dataset was used for fine-tuning the overall systems to gain the ability of emotion generation. All speech were resampled to 16kHz. The phoneme-level alignments were extracted with the MFA\footnote{https://mfa-models.readthedocs.io/en/latest/}.

\subsection{Experiment Set }
\label{sec:exp set}
We conducted the model training on eight NVIDIA V100 GPUs, using the Adam optimizer with $\beta_1$ = 0.8, $\beta_2$ = 0.99. We used the pretrained Emotion2Vec+ Large \footnote{https://github.com/ddlBoJack/emotion2vec} and trained the MPEE on MEAD-TTS for 100 epochs. We first pretrained the AM model on the LibriTTS dataset for 500k steps to ensure its capability in mel-spectrogram generation. The warmup step for the prosody encoder codebook was set to 40k. Subsequently, we fine-tuned the AM model on the MEAD-TTS dataset for 50 epochs. Additionally, the prosody predictor was trained on MEAD-TTS for 50 epochs.

The content encoder is conformer-based\cite{conformer} model with 5 layers and 512 embedding dimensions. The speaker encoder is similar to the Ecapa-TDNN\cite{ecapa}. The duration predictor is a 5-layer Conv-1D with ReLU activation and layer normalization with a hidden size of 512. The prosody predictor is an LLM-like model that contains 8 Transformer layers with 8 attention heads, 768 embedding dimensions. The diffusion decoder is the U-net\cite{unet} architecture with 512 hidden size. We  utilize the pretrained HIFIGAN\cite{hifi}\footnote{https://huggingface.co/speechbrain/tts-hifigan-libritts-16kHz} to  convert the generated mel-spectrogram to waveform which was trained on 16kHz LibriTTS.

\subsection{Evaluation Metrics}
\label{sec:metrics}
We evaluate the speech quality and similarity by both objective metrics and subjective metrics. The objective metrics include the emotion accuracy (EACC) and Words Error Rate(WER). The EACC metric is used to evaluate the emotion consistency between target speech and emotion prompt. We train a speech emotion recognition(SER) model based on the pretrained wav2vec2.0\cite{wav2vec} for EACC evaluation. The WER metric is used to evaluate the content consistency between target and content prompt which is evaluated by Whisper\cite{whisper}-Large 4\footnote{https://github.com/openai/whisper}.

For subjective evaluations, we conduct 5-scale Mean Opinion Score (MOS), Emotion Similarity Mean Opinion Score (ESMOS) and Speaker Similarity Mean Opinion Score (SSMOS) test between MPE-TTS and the baseline models. All scores are reported with a 95\% confidence interval.

\subsection{Experiment Results and Analysis}
\label{sec:results}

We compare our system with other methods in three aspects: speech-based zero-shot generation, image-based zero-shot generation and text-based zero-shot generation.

For speech-based zero-shot generation, we compare our results with those of  GenerSpeech\cite{generspeech} and Meta-StyleSpeech\cite{meta} which were evaluated based on official implementations.

For image-based and text-based zero-shot generation, we select the MM-TTS\cite{MMTTS}, which is the only multi-modal ZS-TTS system, as the baseline model. However, the absence of an official open-source project for MM-TTS poses challenges for a fair comparison with our system. Therefore, we reproduced the experimental results based on the settings described in the original paper.

As shown in Table \ref{tabel:Table1}, among various models that use speech as the prompt, our method achieves competitive performance in both subjective and objective indicators, particularly in terms of quality and emotion consistency. Table \ref{tabel:Table2} and \ref{tabel:Table3} demonstrate that, in the text and image modalities, our method significantly outperforms the baseline model MM-TTS, largely due to the more refined modeling of emotion features in our system. When the emotion prompt is image or text, the SSMOS of MPE-TTS also performs excellently, indicating the effectiveness of the proposed disentangling strategy. Users can input any combination of timbre and emotion to customize the desired speech.

Ablation experiments demonstrate that the proposed MPEE bring more accurate extraction of emotion information compared to the method of MM-TTS. Besides, ECL effectively help the prosody predictor to predict the target prosody which is highly related to emotion. Particularly, the ECL plays a more significant role when using the text and image prompts. Our more sophisticated design enables the system to focus more on emotion information, resulting in better emotion consistency.

\section{Discussion}
Aiming at improving the Zero-Shot generation capabilities of ZS-TTS systems, we propose a multi-modal prompt emotion encoder based on pre-trained Emotion2Vec, which enables our TTS system to support multi-modal prompt input. We introduce an LLM-like prosody predictor to predict emotion-wise prosody, enhancing the emotional alignment between the prompt input and the generated speech. The targeted design for different speech attributes enables our method generate the realistic and rhythmic speech of unseen speakers. In summary, MPE-TTS has achieved flexible and high-quality zero-shot speech synthesis.

This paper acknowledges its limitations. Owing to constraints in emotional data and computational resources, the proposed MPE-TTS has significant potential for improvement in scalability. We believe that larger training data and increased model sizes could enhance zero-shot generation quality.  What’s more, our research has focused solely on enhancing the flexibility of ZS-TTS and has not yet explored the impact of multi-modal fusion features on zero-shot generation.

\bibliographystyle{IEEEtran}
\bibliography{mybib}

\end{document}